\documentclass[aps,prb,twocolumn,amsfonts,amssymb,superscriptaddress,floatfix,longbibliography,10pt]{revtex4-2}
\usepackage{graphicx}
\usepackage{dcolumn}
\usepackage{bm}
\usepackage{wrapfig}
\usepackage{bbm,bbold}
\usepackage{color}
\usepackage{xcolor, soul}
\sethlcolor{green}
\usepackage{amsmath}
\usepackage{mdframed}
\usepackage[normalem]{ulem}
\usepackage{orcidlink}

\usepackage{mathrsfs}   
\usepackage[none]{hyphenat}
\usepackage{cancel}
\usepackage{float}
\usepackage{physics}
\usepackage{comment}
\graphicspath{ {./figures/} }

\usepackage{hyperref}
\hypersetup{colorlinks=true,citecolor=blue,linkcolor=blue,urlcolor=blue}

\begin{document}

\title{Robustness of Magic in the quantum Ising chain via Quantum Monte Carlo tomography}

\author{Hari Timsina~\orcidlink{0009-0003-1044-9856}}
\email{htimsina@sissa.it}
\affiliation{\href{https://ror.org/004fze387}{International School for Advanced Studies (SISSA)}, Via Bonomea, 265, 34136 Trieste, Italy}

\author{Yi-Ming Ding~\orcidlink{0009-0009-9128-9850}}
\email{dingyiming@westlake.edu.cn}
\affiliation{State Key Laboratory of Surface Physics and Department of Physics, Fudan University, Shanghai 200438, China}
\affiliation{Department of Physics, School of Science and Research Center for Industries of the Future, Westlake University, Hangzhou 310030,  China}
\affiliation{Institute of Natural Sciences, Westlake Institute for Advanced Study, Hangzhou 310024, China}

\author{Emanuele Tirrito~\orcidlink{0000-0001-7067-1203}}
\email{etirrito@ictp.it}
\affiliation{\href{https://ror.org/009gyvm78}{The Abdus Salam International Centre for Theoretical Physics (ICTP)}, Strada Costiera 11, 34151 Trieste, Italy}
\affiliation{Dipartimento di Fisica ``E. Pancini", Universit\`a di Napoli ``Federico II'', Monte S. Angelo, 80126 Napoli, Italy}

\author{ 
Poetri Sonya Tarabunga~\orcidlink{0000-0001-8079-9040}}
\email{poetri.tarabunga@tum.de}
\affiliation{Technical University of Munich, TUM School of Natural Sciences,
Physics Department, 85748 Garching, Germany}
\affiliation{Munich Center for Quantum Science and Technology (MCQST),
Schellingstr. 4, 80799 M\"unchen, Germany}

\author{Bin-Bin Mao}
\email{maobinbin@uor.edu.cn}
\affiliation{School of Foundational Education, University of Health and Rehabilitation Sciences, Qingdao 266000, China}

\author{Mario Collura~\orcidlink{0000-0003-2615-8140}}
\affiliation{\href{https://ror.org/004fze387}{International School for Advanced Studies (SISSA)}, Via Bonomea, 265, 34136 Trieste, Italy}
\affiliation{INFN, Sezione di Trieste, Via Valerio 2, 34127 Trieste, Italy}

\author{Zheng Yan}
\email{zhengyan@westlake.edu.cn}
\affiliation{Department of Physics, School of Science and Research Center for Industries of the Future, Westlake University, Hangzhou 310030,  China}
\affiliation{Institute of Natural Sciences, Westlake Institute for Advanced Study, Hangzhou 310024, China}

\author{Marcello Dalmonte~\orcidlink{0000-0001-5338-4181}}
\affiliation{\href{https://ror.org/009gyvm78}{The Abdus Salam International Centre for Theoretical Physics (ICTP)}, Strada Costiera 11, 34151 Trieste, Italy}

\date{\today}

\begin{abstract}

We study the behavior of magic as a bipartite correlation in the quantum Ising chain across its quantum phase transition, and at finite temperature. In order to quantify the magic of partitions rigorously, we formulate a hybrid scheme that combines stochastic sampling of reduced density matrices via quantum Monte Carlo, with state-of-the-art estimators for the robustness of magic - a {\it bona fide} measure of magic for mixed states. This allows us to compute the mutual robustness of magic for partitions up to 8 sites, embedded into a much larger system. We show how mutual robustness is directly related to critical behaviors: at the critical point, it displays a power law decay as a function of the distance between partitions, whose exponent is related to the partition size. Once finite temperature is included, mutual magic retains its low temperature value up to an effective critical temperature, whose dependence on size is also algebraic. This suggests that magic, differently from entanglement, does not necessarily undergo a sudden death.

\end{abstract}

\maketitle

\section{\label{sec:intro} Introduction}
There is presently considerable interest in understanding the complexity properties of quantum many-body states from the viewpoint of quantum error correction~\cite{knill1997theory,terhal2015quantum,lidar2013quantum}. Such interest has been triggered by a combination of facts, including new insights from quantum information theory and atomic physics experiments demonstrating basic elements of error correction with tens of logical qubits~\cite{cory1998experimental,chiaverini2004realization,pittman2005demonstration,ofek2016extending,ryan2021realization}. The complexity of realizing a quantum state in a fault-tolerant manner is directly related to quantifying the operations for which error correction itself is most challenging (e.g., most time-consuming). In the context of stabilizer codes~\cite{gottesman1997stabilizer,aaronson2004improved}, encompassing a large portion of known error correction architectures, these operations have been referred to as magic gates, and the corresponding quantum resource as magic - or non-stabilizerness~\cite{bravyi2005universal,howard2014contextuality,campbell2017roads,liu2022manybody}. Quantifying the magic of a state will thus inform on the difficulty of realizing or approximating the state itself, utilizing a fault-tolerant functioning quantum computer~\cite{bravyi2016improved,bravyi2016trading,wang2020efficiently,RevModPhys.91.025001,gottesman1999demonstrating,bluvstein2025architect,tarabunga2024nonstabilizernessmonotone}. 

For the specific case of pure states, there are now both concepts (measures and witnesses), tools, and theories that have allowed us to scrutinize the relation between magic and physical phenomena. From the conceptual viewpoint, a breakthrough was the introduction of stabilizer R\'enyi entropies (SREs)~\cite{leone2022stabilizer,Haug2023stabilizerentropies,PhysRevA.110.L040403}, which allowed for casting the problem of quantifying magic from minimization on a non-convex space to the computation of expectation values. This result triggered immediate interest in the computational physics community, based on developing efficient algorithms for computing SRE. These algorithms, relying on the probabilistic nature of SREs and the Matrix Product State (MPS) representation of quantum states, include the Pauli-Markov chain~\cite{tarabunga2023many}, Perfect sampling~\cite{lami2023nonstabilizerness,Haug2023stabilizerentropies}, Pauli MPS technique~\cite{tarabunga2024nonstabilizerness} and multi-replica methods~\cite{haug2023quantifying}, and provide efficient and well-behaved estimates of magic for the pure states. 
Together with field theory insights, it is by now established that magic is deeply connected to quantum critical behavior~\cite{ding2025evaluating,hoshino2025stabilizer,hoshino2025stabilizerrenyientropyencode}. 

The understanding of the magic of partitions - that is, of mixed states density matrices - is, instead, considerably less mature. This is mostly due to the fact that, in the case of qubits, SREs and other measures cast as expectation values of operators are not applicable. For qubit mixed states, magic measures do exist, such as robustness of magic~\cite{howard2017application} and relative entropy of magic~\cite{veitch2014resource,liu2022manybody}, but are computationally expensive to compute exactly. So far, only studies of single-site partitions in solvable models have been pursued: while these results have reported an interesting connection between magic and critical behavior, it is hard to understand the role of the former given the strong correspondence between traditional correlation functions and magic for that specific scenario. In particular, we presently lack a clear understanding of {\it mutual magic} - that is, the magic between two separate partitions -, which could complement the widespread and key relevance of mutual information, its parallel in the context of entanglement. 

In this work, we study the mutual magic between partitions in the quantum Ising model, both in the ground state and at finite temperature, focusing on the robustness of magic. To compute the latter, we introduce a hybrid method that is based on a combination of quantum Monte Carlo sampling and tomography, with an optimized scheme for robustness estimation. This allows us to consider modest partition sizes embedded in large systems, with no restriction on the partition positioning, and no assumption on the structure of the state (e.g., Gaussian, stabilizer, etc.). 

The structure of the paper is as follows. In Sec.~\ref{sec:preliminaries}, we introduce the magic and robustness of magic, and formulate the hybrid scheme used to compute the latter at scale. In Sec.~\ref{sec:resultsdiscussion}, we present our results. After a warm-up investigation of the full state robustness of magic, which is already sensitive to critical behavior in a qualitative manner, we focus on mutual magic correlations. The latter shows strong features of quantum criticality, decaying as a power law of the distance between partitions for various partition sizes. A precise connection to critical exponents is, however, hard to determine, a feature we attribute to the complex minimization nature of the robustness of magic. We then discuss how magic is resilient to finite temperature effects, in particular at critical points. Finally, we draw our conclusions in Sec.~\ref{sec:conclusion}.

\section{\label{sec:preliminaries} Preliminaries} 
The Robustness of Magic (RoM) measures nonstabilizerness, a concept rooted in the Clifford formalism of quantum error correction. This Section is devoted to reviewing basic concepts in a self-contained manner. 
In Sec.~\ref{subsec:clifford}, we define the Clifford group as the normalizer of the Pauli group. We also introduce stabilizer states and explain why nonstabilizer operations are necessary for universal quantum computation.
In Sec.~\ref{subsec:Rom}, we define RoM as a quantifier of nonstabilizerness, along with its variants: the log-free RoM and mutual-RoM. The computation of RoM requires the partition density matrix, and Sec.~\ref{subsec:numerical} discusses different numerical methods for this calculation we employ here.

\subsection{\label{subsec:clifford} Clifford group and stabilizer states}
Let $X, Y, Z$ be the Pauli operators, and together with the Identity operator $I$, we call $\mathcal{P}_1=\{I, X, Y, Z\}$ the Pauli string for a qubit. The generalization of those to $n$-qubits $\mathcal{P}_n = \mathcal{P}_1^{\otimes n}$ form the Pauli-$n$ string, the members of the Pauli-$n$ group with phase $+1$ only.
Next, we define the Clifford group as the normalizer of the Pauli group: 
\begin{equation}
    \text{Cl}_n = \{U:\, UPU^\dagger \in \mathcal{P}_n,\, \forall P\in \mathcal{P}_n \}.
\end{equation}
The Clifford gates can be constructed by the generators consisting of single-qubit Hadamard (H), and phase $(\pi/2)$ S, and two-qubit controlled-NOT (CNOT) gates~\cite{nielsen2010quantum}. 
The Clifford operations can create an arbitrary amount of entanglement between qubits.
The application of $n$-Clifford operators on the computational basis state $|00\cdots 0\rangle$ gives rise to the class of states known as the stabilizer states~\cite{gottesman1997stabilizer}, denoted as $|s_n\rangle = \text{Cl}_n|0\rangle^{\otimes n}$. 
We call $\{\mathcal{S}_n\}$ the set of all pure $n$-qubit stabilizer states, total number of such states is given by~\cite{aaronson2004improved,garcia2017geometry}
$$|\mathcal{S}_n| = 2^n \prod_{k=1}^{n} \left(2^{k} +1\right) \approx 2^{O(n^2)}.$$
The convex hull of the stabilizer states,  defined as a probabilistic mixture of pure stabilizer states $\sigma_i=|s_n\rangle\langle s_n|$
\begin{equation}\label{eq:stabn}
     {\rm STAB}_n = \left\{\sum_i x_i\sigma_i : \sigma_i\in \mathcal{S}_n,\, x_i\geq 0,\,\sum_i x_i=1 \right\},
\end{equation}
includes all mixed stabilizer states and constitutes a stabilizer polytope.
The stabilizer protocol, accompanied by only Clifford operations, Pauli measurements, and classical control, can be classically simulated in polynomial time via the Gottesman-Knill protocol~\cite{gottesman1998heisenberg}; the above are considered free operations.

Clifford gates are thus not universal and require an additional candidate for the universal quantum computation, which must include the states not accounted for by Eq.~\eqref{eq:stabn}.
This is fulfilled by the T-gate $(\rm diag(1,e^{i\pi/4}))$ operation, resulting in the new resource of quantum information, called nonstabilizerness (or magic)~\cite{bravyi2005universal}. 
Magic thus represents a key element to access quantum advantage: due to this, its physical consequences on many-body states have recently attracted considerable interest. Most studies in this area have focused on SREs because they are efficiently computable and scalable, and provide the experimental realization~\cite{oliviero2022measuring,niroula2024phase,bluvstein2024logical,turkeshi2024coherent,rodriguez2024experimental}. In particular, they are useful to study the phase behavior in (quasi)1D systems and are able to capture the critical behavior for the ground state of the systems~\cite{Tarabunga2024criticalbehaviorsof,tarabunga2023many,frau2024nonstabilizerness,falcao2025nonstabilizerness,tarabunga2023magic,PhysRevB.111.054301,viscardi2025interplay,robin2025stabilizer}. Further, they are also studied in the context of neural quantum states~\cite{mello2025retrieving,sinibaldi2025nonstabilizerness,spriggs2025quantum}. Unfortunately, SREs lack contact with resource theory for the mixed states, i.e., the states that are only representable by the density matrices. This calls for the investigation of critical behavior for those systems with a well-defined measure of magic for the mixed states, also. To this end, we define the Robustness of magic.

\subsection{\label{subsec:Rom} Robustness of Magic}
Robustness of entanglement was put forward in Ref.~\cite{vidal1999robustness} by defining an entanglement monotone as the minimum amount of separable noise that makes an entangled state separable.  
The entangled state in a finite-dimensional Hilbert space of the composite system can be decomposed as $\rho = (1+p)\rho_{+} - p\rho_{-}$ in terms of the two separable states $\{\rho_+, \rho_-\}$ and a finite real coefficient $p\geq 0$~\cite{vidal1999robustness,stiner2003generalized}.
The same concept can be applied to the expansion of the generic quantum state in terms of the pair of stabilizer states  $\{\sigma_+, \sigma_-\}$. It is clear that for the stabilizer state itself, the coefficient vanishes $p=0$.
Generalizing this, we can define a magic monotone called robustness of magic as the minimum overlap of a nonstabilizer state with the members of a stabilizer polytope~\cite{howard2017application,heinrich2019robustness,seddon2019quantifying}. In other words, the robustness of magic $\mathcal{R}$ quantifies the minimum weight of the combination of stabilizer states that can reproduce a given quantum state. Mathematically, it is defined by writing $\rho$ as the stabilizer pseudomixtures
\begin{equation} \label{eq:rom}
    \mathcal{R}(\rho) = \min_{\mathbf{x}\in \mathbb{R}^{|\mathcal{S}_n|}} \left\lbrace \|\mathbf{x}\|_1 : \quad \rho = \sum_{i=1}^{|\mathcal{S}_n|} x_i\sigma_i,\, \sigma_i\in\mathcal{S}_n \right\rbrace.
\end{equation}
Since $\sum_i x_i=1$ holds, the robustness of a given state measures the amount of negativity in the aﬃne combination through the $l_1$-norm $\|\mathbf{x}\|_1=\sum_i |x_i|=1+2\sum_{i,x_i\leq 0}|x_i|$ of the quasi probability distribution $\{\mathbf{x}\}$.
The problem of Eq.~\eqref{eq:rom} can be recast into the convex minimization over the solutions of systems of linear equations in $4^n$-dimensional space for the $|\mathcal{S}_n|$ variables
\begin{equation} \label{eq:rom1}
    \mathcal{R}(\rho) = \min_{\mathbf{x}\in \mathbb{R}^{|\mathcal{S}_n|}} \left\lbrace \|\mathbf{x}\|_1 \quad \text{subject to} \quad \mathbf{A}_n\mathbf{x} = \mathbf{b} \right\rbrace,
\end{equation}
where $A_{ji} = {\rm Tr}(\sigma_i P_j), \, b_j = {\rm Tr}(P_j \rho)$ and $P_j\in\mathcal{P}_n\, (1\leq j \leq 4^n)$~\cite{howard2017application}.
The definition of Eq.~\eqref{eq:rom1}  guarantees a feasible and bounded solution; therefore, there exists a strong duality and defines the dual version of the RoM in terms of the inequality-based maximization problem
\begin{equation} \label{eq:romdual}
    \mathcal{R}(\rho) = \max_{\mathbf{y}\in \mathbb{R}^{4^n}}\{ \mathbf{b}^\top\mathbf{y}:\,\, -\mathbf{1}\leq \mathbf{A}_n^\top\mathbf{y} \leq \mathbf{1} \},
\end{equation}
where $\mathbf{1}$ is a length-$|\mathcal{S}_n|$ vector with all the elements given by unity~\cite{howard2017application,hamaguchi2024handbook} and $\mathbf{y}$ is the dual variable of $\mathbf{x}$. 

The RoM is a good measure of magic in the context of resource theory as it satisfies all the required properties for arbitrary quantum states~\cite{howard2017application}.
In particular, it is faithful in the sense that $\mathcal{R}(\rho) \geq 1$; the equality holds if and only if $\rho$ is a stabilizer state. 
RoM is nonincreasing under stabilizer operations, and it follows the submultiplicative behavior $\mathcal{R}(\rho_1\otimes \rho_2) \leq \mathcal{R}(\rho_1)\times \mathcal{R}(\rho_2)$.  
It is a magic monotone, for all trace-preserving stabilizer channels $\mathcal{E}$ we have $\mathcal{R}(\mathcal{E}(\rho)) \leq \mathcal{R}(\rho)$, and also the average robustness is non-increasing under a trace non-increasing map.
The RoM $(\mathcal{R}(\rho))$ characterizes the overhead of classically simulating a quantum circuit that uses a magic state $\rho$ as an ancilla. In quasiprobability Monte Carlo simulation, the simulation cost, measured by the number of samples required to achieve a fixed accuracy, scales as $O(\mathcal{R}(\rho)^2)$, which makes robustness a direct quantifier of classical simulation complexity~\cite{heinrich2019robustness}.

The numerical evidence suggests that the RoM of a quantum state depends extensively on the number of qubits and its value grows subexponentially for copies of $|H\rangle$ and $|T\rangle$ states~\cite{heinrich2019robustness}. The linearized-robustness $\mathcal{R}(|\psi\rangle^{\otimes n})^{1/n},\, \psi=H, T$ for those states is seen to converge towards a value lower than that of the single-qubit value. We can also take the logarithm of RoM and define the associated measure of magic called  Log-free robustness of magic (LRoM)
\begin{equation}
    \rm LR(\rho) = \log_2 \mathcal{R}(\rho).
\end{equation}
The monotonic property is preserved, but submultiplicative transforms to the subadditive property  $\rm LR(\rho_1\otimes \rho_2) \leq LR(\rho_1) + LR(\rho_2)$ and the faithfulness becomes $\rm LR(\rho)=0$ for the stabilizer states.

RoM can also serve as an upper bound to many of the other measurable quantities. Those bounds are valuable since numerical methods are limited to modest numbers of qubits, whereas the bounding relations will hold for any number of qubits.
The RoM for the $n$-qubit state $\rho$ is bounded as  
$\mathcal{D}(\rho) \leq \mathcal{R}(\rho)$, where $\mathcal{D}(\rho) = 2^{-n}\sum_{P \in \mathcal{P}_n} |{\rm Tr}(P\rho)|$ is the average of absolute overlaps with the Pauli strings
\footnote{$\mathcal{D}(\rho)$ satisfies the following properties: (i) Convexity: $\mathcal{D}\left(\sum_k p_k\rho_k\right) \leq \sum_k |p_k|\mathcal{D}(\rho_k)$,  (ii) Magic witness: if $\mathcal{D}(\rho)>1$ then $\rho$ is a nonstabilizer state, (iii) Multiplicativity: $\mathcal{D}(\rho_1\otimes\rho_2) = \mathcal{D}(\rho_1)\mathcal{D}(\rho_2)$.}.  
RoM can be related to the SREs~\cite{leone2022stabilizer}, which are well-defined measures of magic for pure states $\ket{\psi}$ and efficiently calculated numerically for large systems using tensor networks and quantum Monte Carlo, defined as 
\begin{equation} \label{eq:sre}
    M_\alpha(\rho=|\psi\rangle\langle\psi|)=\frac{1}{1-\alpha}\ln A_\alpha(\rho),
\end{equation}
where $A_\alpha(\rho)$ is the $\alpha$-moment of the Pauli spectrum~\cite{haug2023efficient}
\begin{equation}
    A_\alpha(\rho)=2^{-n}\sum_{P\in\mathcal{P}_n} \vert\text{tr}(\rho P)\vert^{2\alpha}\,.
\end{equation}
The SREs can be generalized to mixed states, which, while they are not proper measures, serve as magic witnesses~\cite{haug2025efficientwitnessingtestingmagic}. The SRE witness is defined as
\begin{equation}\label{eq:srewitness}
    \mathcal{W}_\alpha(\rho)= \frac{1}{1-\alpha}\ln A_\alpha(\rho) - \frac{1-2\alpha}{1-\alpha}S_2(\rho)\,,
\end{equation}
where $S_2(\rho)=-\ln\text{tr}(\rho^2)$ is the 2-R\'enyi entropy. 
The LRoM relates to the SRE witness as $\mathcal{W}_\alpha\leq 2\text{LR}$~\cite{haug2025efficientwitnessingtestingmagic}. 

In quantum computation and many-body systems, it's not just the properties of individual states that matter; the real insight often comes from how different parts of the system relate to each other, through correlations, entanglement, and other shared features.
A useful way to capture this kind of shared behavior is through mutual information between the different subsystems.
It gives us a measure of how much knowing something about one part tells us about another, and it's especially helpful when studying systems that aren't in a pure quantum state. They are used to understand entanglement patterns, detect phase transitions, or track how information spreads across a system. The mutual SRE~\cite{tarabunga2023many} is also defined as the study of the magic spreading in the quantum spin systems accompanied by the presence of entanglement. Like the SRE, it has been related to physical phenomena~\cite{tarabunga2023many,tarabunga2024magictransition,frau2024stabilizer,lopez2024exact,hoshino2025stabilizer,korbany2025long,tarabunga2025efficientmutualmagicmagic,PRXQuantum.6.020324}.
With this motivation, we will compute magic for different subsystems in a single state to distinguish local magic from nonlocal one. If, for instance, $\rm LR(\rho_{12}) > LR(\rho_1\otimes \rho_2 )$, we will then conclude that magic is stored nonlocally in the joint system $\rho_{12}$, which is not found in each of the individual subsystems. To this end, we define the mutual robustness (MLRoM) in terms of LRoM as the difference 
\begin{equation} \label{eq:mutualrom}
    \rm MR(\rho_{12}) = LR(\rho_{12}) - LR(\rho_{1} \otimes\rho_{2}).
\end{equation}
This definition ensures that $\rm MR\geq 0$ for arbitrary bipartitions of the correlated systems~\cite{sarkar2020characterization}, and the equality is satisfied for the product states. Due to the differences, MLRoM is free of any cutoff of a field theory by canceling the boundary terms, which give rise to universal physical properties.

Before continuing further, it is worth pointing out that, for single-site partitions, a recent work~\cite{sarkar2020characterization} has already investigated the behavior of magic for single sites, and between pairs of sites at a distance. In the context of the Ising chain, Ref.~\cite{sarkar2020characterization} has already pointed out how magic can be related to quantum critical behavior: however, the limitation in terms of partition sizes makes those results of not easy interpretation, as quantum resources are essentially only sensitive to two-body correlation functions for such tiny partitions~\footnote{This is simply because one can perform full state tomography from two-body correlations in a two-site partition}. Our goal here is to explore regimes where this is not possible - that is, exploring the behavior of magic in large systems, and sizeable partitions whose properties cannot be generically ascertained by two-body correlations.

\subsection{\label{subsec:numerical} Numerical Techniques}
\subsubsection{\label{subsubsec:dmcalc} Calculation of density matrix} 
\begin{figure}[ht]
    \centering
    \includegraphics[width=1\linewidth]{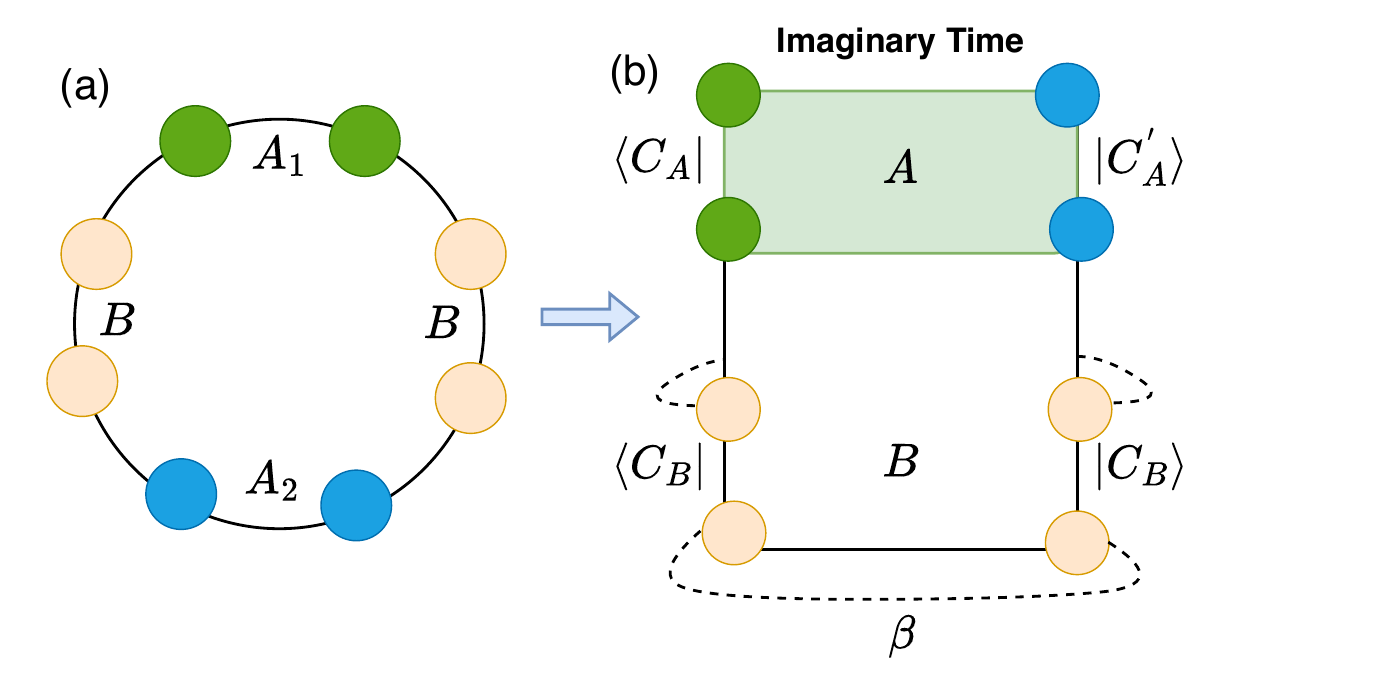}
    \caption{(a) 1D spin chain with periodic boundary conditions. The system is transferred to (b) for the RDM calculation through QMC using SSE. 
    We calculate the reduced density matrix by tracing out the $B$ part, leaving two disconnected subsystems $A_1$ and $A_2$.
    }
    \label{fig:1dchainpbc}
\end{figure}

Calculating the Robustness of Magic (RoM) requires knowledge of the density matrix of the system under interest.
For physical quantum systems, the density matrix can be obtained either exactly through exact diagonalization (ED) or approximated using various numerical techniques such as Quantum Monte Carlo (QMC), density matrix renormalization group (DMRG), and Tensor Networks (TN), among others. 
In this work, ED is used to study the full system's RoM; meanwhile, we use QMC to sample the reduced density matrix (RDM) of the subsystem by tracing the environment degrees of freedom~\cite{mao2025sampling,mao2025detecting} and then do the RoM calculations. 
To achieve this, we consider a tripartite division of the 1D system with periodic boundary condition (PBC) into two disconnected subsystems $A_1$, $A_2$, and the remaining environment $B$ (as illustrated in Fig.~\ref{fig:1dchainpbc}). 
By tracing out the environmental degrees of freedom, we obtain the RDM for the combined small subsystem $A_1+A_2\equiv A$.  
The RDM element $\langle C_{A}| \rho_A |C'_A\rangle$ can be written as
\begin{equation} \label{eq:rdmeq}
    \langle C_{A}| \rho_A |C'_A\rangle\propto \frac{\sum_{\{C_B\}} \langle C_{A},C_B | e^{-\beta H}| C'_{A},C_B\rangle}{\sum_{\{C_A,C'_A,C_B\}} \langle C_{A},C_B | e^{-\beta H}| C'_{A},C_B\rangle},
\end{equation}
where $C_{A}, C'_{A}$ are the configurations of bra and ket corresponding to the RDM, and $\{C_B\}$ means all the configurations of environment $B$, $\beta=1/T$ is the inverse temperature ($\beta\rightarrow \infty$), and $H$ is the Hamiltonian of the full system containing $A$ and $B$. 
Within QMC, we used the stochastic series expansion (SSE)~\cite{Sandviksusc1991, Sandvik1999, yan2019sweeping, yan2022global,sandvik2019stochastic,sandvik2010computational,tarabunga2025bell} to compute the RDM, where the imaginary-time boundaries are modified such that the boundary of imaginary time in the region $A$ is open but the periodic boundary condition is kept for the environment $B$. 
For calculating the RDM of a single partition (say $A_1$), we attach another partition $(A_2)$ to the environment and proceed accordingly.
The Eq.(\ref{eq:rdmeq}) means the element of RDM $\langle C_{A}| \rho_A |C'_A\rangle$ is proportional to the frequency of the QMC samplings that bra is $C_A$ and ket is $C'_A$~\cite{mao2025sampling}, that is, $\langle C_{A}| \rho_A |C'_A\rangle\propto N_{C_A,C'_A}/M$ where $M$ is the total sampling number and $N_{C_A,C'_A}$ is the sampling number in which bra is $C_A$ and ket is $C'_A$.
The detail of the SSE QMC method is described in Appendix~\ref{appendix:rdm}.

\subsubsection{\label{subsubsec:romcalc} Calculation of RoM}
The calculation of RoM, as defined in Eqs.~\eqref{eq:rom} and~\eqref{eq:rom1}, is the linear programming (LP) problem.
Although the time complexity of LPs is linear in the product of the number of constraints and variables, these numbers themselves grow superexponentially with the number of qubits because of the size of the stabilizer group.
Therefore, the computation of RoM is challenging in terms of time and memory and is limited to a few qubits. 
Following Ref.~\cite{howard2017application}, we have the exact $\mathbf{A}_n$-matrix used in Eq.~\eqref{eq:rom1} up to $n=5$ qubits, and we can compute the RoM naively for $n\leq 5$ for any arbitrary state through convex optimization~\cite{howard2017application,aaronson2004improved} 
For $n\leq 8$, we can use the iterative procedure named Column Generation (CG) method developed in Ref.~\cite{hamaguchi2024handbook}. The CG method is capable of yielding the exact results and works for pure- as well as mixed-state density matrices. 
Even for the systems with some local or global symmetries, or the states obtained from the tensor products of single-qubit states, the approximate calculation can only be extended up to a few tens of qubits (up to 26/24 copies of $H/T$ state)~\cite{heinrich2019robustness}. 
Since we work on the quantum states governed by a fully dense density matrix, which typically does not possess any symmetry to reduce the complexity, we can only employ an exact method and CG technique to compute the RoM.
More details about these methods are presented in Appendix~\ref{appendix_CG}.

\section{\label{sec:resultsdiscussion} Results and Discussions}
We now provide the numerical results of the RoM for the particular many-body system of a physical model, the one-dimensional transverse field Ising model (1D TFIM).
Consider $L$ spins in a 1D lattice with nearest neighbor interaction governed by the TFIM Hamiltonian~\cite{pfeuty1970onedimensional}
\begin{equation} \label{eq:1dtfim}
    H = -J \sum_{i=1}^L \sigma_{i}^z \sigma_{i+1}^z - h \sum_{i=1}^L \sigma_{i}^x,
\end{equation}
where $J>0$ and $h>0$ are ferromagnetic exchange interaction and external field strength, respectively, and $\sigma^\alpha (\alpha=x,y,z)$ are the Pauli spin matrices. 
The model in Eq. \eqref{eq:1dtfim} is integrable by Jordan-Wigner transformation and possesses a quantum critical point (QCP) $h/J=1$ \cite{pfeuty1970onedimensional,sachdev2011quantum,bigan2024quantumising}. 
For simplicity, we put $J=1$ in Eq. \eqref{eq:1dtfim} and we assume the periodic boundary condition (PBC) where the last and first spins are connected in the ring $(\sigma_{L+1}=\sigma_1)$ as shown in Fig.~\ref{fig:1dchainpbc}. 
The eigenstates are stabilizer states only in specific cases of  $h=0$ (ferromagnetic ground state) and $h\rightarrow\infty$ (fully polarized state) for an arbitrary $\beta$. 
The nonstabilizerness will be present for all other finite field strengths except for $\beta=0$.
In the following, all calculations of RoM are performed for the thermal density matrix with the related measures LRoM and MLRoM.

\subsection{\label{subsec:fullresult} Full system RoM}

As a methodological first step, we begin by computing the RoM for the thermal density matrix of the 1D TFIM without partitioning the system. This case is also interesting since the full state SRE of conformally critical ground states has been recently shown to be universal~\cite{hoshino2025stabilizer,hoshino2025stabilizerrenyientropyencode}, and determined by the ground-state degeneracy of the boundary conformal field theory imposed by the basis whose magic we are interested in.

\begin{figure}[ht] 
    \centering
    \includegraphics[width=8cm]{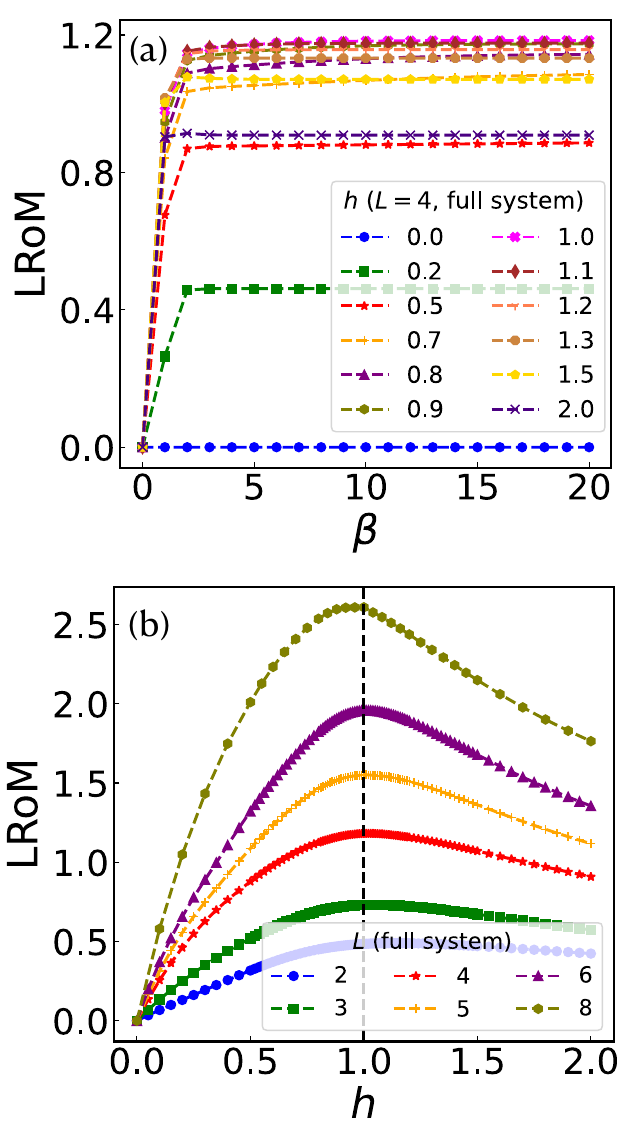}
    \caption{\label{fig:romfull} 
    (a) LRoM for the full system of the 1D TFIM with periodic boundary conditions (PBC) as a function of $\beta$ for different values of $h$. The LRoM saturates and forms a stable plateau after a finite $\beta$. 
    (b) LRoM for $L \leq 8$ and ground state approximation $\beta=2L$. 
    }
\end{figure}

While the exact ground-state density matrix cannot be calculated analytically for arbitrary parameter configurations, we approximate it by selecting an appropriate inverse temperature $\beta$. The validity of these approximations is verified by ensuring the convergence of the RoM in the large-$\beta$ limit (effectively the zero-temperature limit).
This is illustrated in Fig.~\ref{fig:romfull}(a), where we plot the variation of LRoM with $\beta$ for different values of the field $h$ in the full system of length $L = 4$.
The stabilizerness of $h=0$ is validated for all $\beta$.
For all other field strengths, starting from the stabilizer state at infinite temperature, LRoM increases with $\beta$, reaches a maximum, and then saturates to a stable plateau. 
This plateau corresponds to the zero-temperature $(T\rightarrow 0)$ limit. This is true because usually when $\beta\sim O(L)$, the imaginary time propagation would leave the system in an approximately pure ground state.
We can argue that the stable value of LRoM at large $\beta$ depends on both parameters $L$ and $h$, and also the zero temperature behavior appears to persist over a finite range of temperatures. 
This indicates that, for the purpose of full-state magic characterization, the system in the stable region can be effectively considered as its ground state. These observations will also hold for the partitioned system, which will be seen in the Sec.~\ref {subsec:reducedresult}.

To study the field dependence of RoM, we approximate the effective GS by the length-dependent value of temperature as large $\beta=2L$. The full-system result is shown in Fig.~\ref{fig:romfull}(b) for the system sizes $L\leq 8$. 
The increase of field $h$ increases LRoM up to the vicinity of QCP and starts to decrease, which separates the ordered and disordered regions. 

\subsection{\label{subsec:reducedresult} RoM for reduced system}
The reduced system considered here is the tripartition system with two disconnected subregions $A_1$ and $A_2$ as shown in Fig.~\ref{fig:1dchainpbc}. We consider symmetric bipartitions of sizes $\ell=2+2$ and $4+4$, corresponding to total subsystem sizes of four and eight, respectively, taken from systems of length $L = \{8, 16, 32, 64, 128\}$. The average and error bars of each measurement are calculated over 30 QMC samples of RDMs. The thermalization and total Monte Carlo steps to obtain each RDM for the $2+2$ subsystem are 50,000 and 500,000, and for $4+4$ are 50,000 and 32,00,000, respectively.

\subsubsection{Field Dependence}
The ground-state behavior of LRoM and MLRoM with the transverse field strength is studied again by considering $\beta=2L$. Further, we present the scaling relation with the size of the system at the QCP for both quantities. The (dis)agreement with the full system magic and also between the two partition sizes is explained.

\begin{figure}[ht]
    \centering
    \includegraphics[width=8cm]{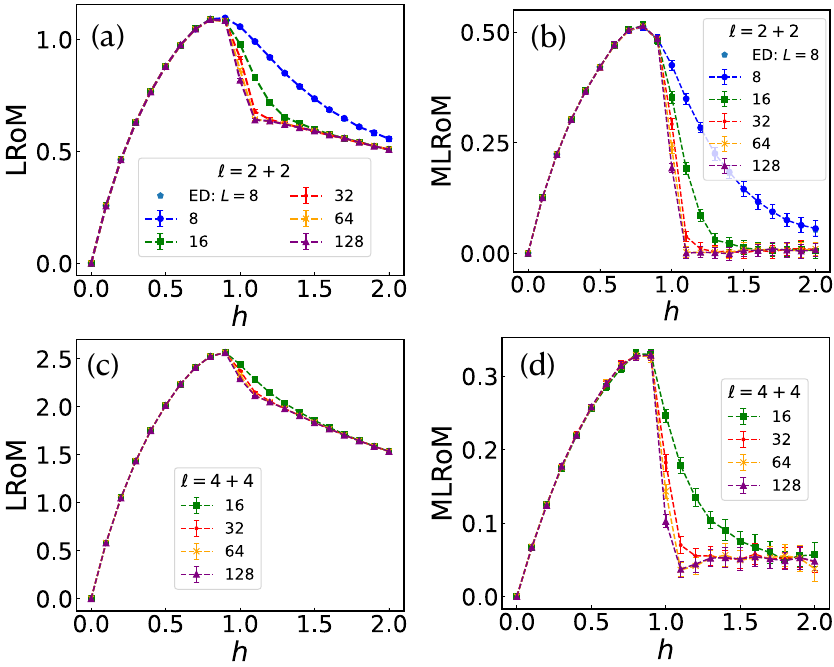}
    \caption{LRoM ((a) size 2+2 and (c) 4+4) and MLRoM ((b) size 2+2 and (d) 4+4) with $h$ for the low temperature state $\beta=2L$.}
    \label{fig:romvsh}
\end{figure}

The zero field reduced system is again a stabilizer indicated with vanishing magic, indicating subsystem of stabilizer is itself a stabilizer. 
As $h$ increases, the magic grows and attains a maximum value after which it decreases towards a finite value. 
Although the peak of LRoM is attained at $h=1$ for the full system magic calculation (see Fig.~\ref{fig:romfull}(b)), the maximum is observed at a lower value than the QCP $(h<1)$ for the tripartition case. 
Up to the peak value at the ordered phase, LRoM is found to be the same and barely depends on the full system size.  
Only after the maximum point, the size dependence becomes evident, as shown in Fig.~\ref{fig:romvsh}(a) for the $2+2$ subsystem and Fig.~\ref{fig:romvsh}(c) for the $4+4$ subsystem.
Therefore, in the critical region, the LRoM is distinguished between different $L$'s, and the value is smaller for larger system sizes.
For $L=8$ and $2+2$ subsystem, we have also presented the benchmark result using RDM between the pure ED and QMC+ED, and they both are in excellent agreement. In fact, RDMs obtained from both methods are similar, as evidenced by their small norm distance as compared with entries in the matrix and nearly identical purities across various samples. 
The corresponding mutual robustness, measured by MLRoM for $2+2$ and $4+4$ partitions, is presented in Figs.~\ref{fig:romvsh}(b) and \ref{fig:romvsh}(d), respectively.
Similar to LRoM, we observe that:
(i) the maximum value is not attained at the critical point, but at a lower value of the field;
(ii) for fields lower than the peak value, MLRoM remains constant and is independent of the total system size;
(iii) size dependence emerges beyond the peak, where the MLRoM is smaller for the larger system compared to the smaller one.
While LRoM approaches a finite value at large field strengths, MLRoM vanishes due to the cancellation of offset terms during the subtraction of the two contributions in Eq.~\eqref{eq:mutualrom}, which is also seen from the figures. 
The errors in LRoM are small and quite difficult to see in the figure, while for the MLRoM, they are clearly visible as the errors are additive during the subtraction of two terms.

In entanglement theory, there are limits to how entanglement can be shared among multiple subsystems. This limitation is known as the monogamy of entanglement~\cite{coffman2000distributed,osborne2006generalmonogamy}, meaning that if two subsystems are highly entangled, they cannot be as strongly entangled as a third subsystem. An analogous concept can be formulated within the framework of magic resource theory. In this context, the \textit{monogamy of magic} provides a similar restriction on the distribution of magic across different parts of a multipartite system. This trade-off relation of sharing magic is captured by the inequality $\text{LR}(\rho_{A_1A_2/B}) \geq \text{LR}(\rho_{A_1/A_2B} \otimes \rho_{A_2/A_1B})$, where $\rho_{A/B}$ denotes the reduced density matrix of the subsystem $A$ taken from the combined system of $A+B$. Our observation reveals that LRoM serves as a meaningful indicator of the monogamy of magic. Therefore, in multipartite quantum systems, it provides a quantitative tool and conceptual framework for understanding the distribution of magic across its partitions.

\begin{figure}[ht]
    \centering
    \includegraphics[width=8cm]{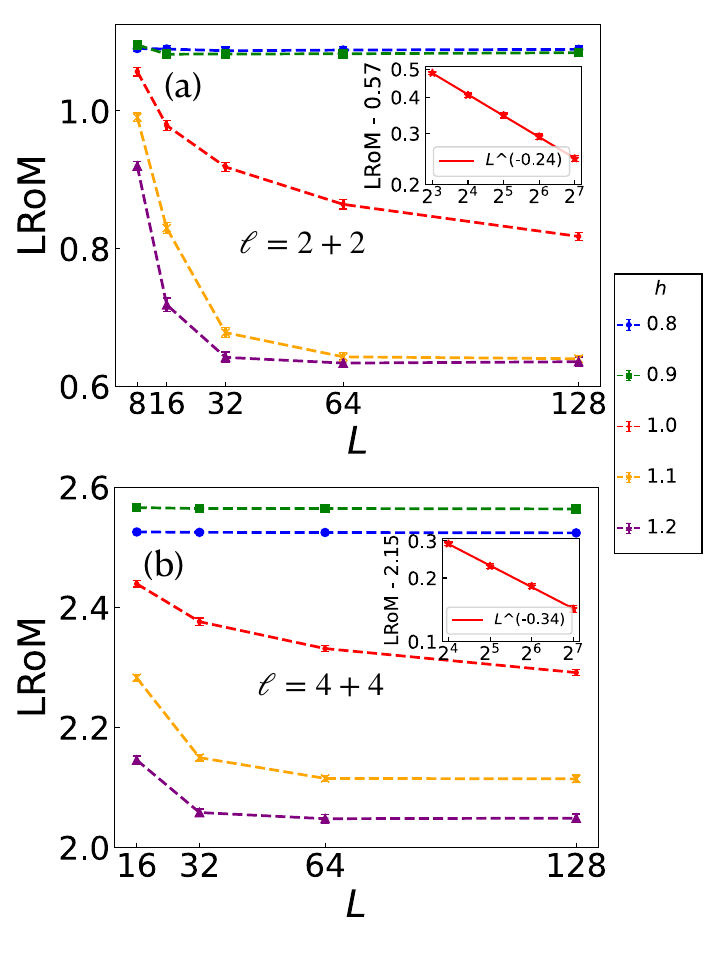}
    \caption{LRoM with system size for different values of $h$ for the partition size of (a) $2+2$ and (b) $4+4$. The power law scaling at $h=1$ is presented in the insets.}
    \label{fig:lrom_size}
\end{figure}

The dependence on total system size for the different field values discussed above is also evident in Fig.~\ref{fig:lrom_size}(a) for 2+2 subsystem and Fig.~\ref{fig:lrom_size}(b) for 4+4 subsystem, where we plot the LRoM as a function of system size for selected values of $h$ near the critical region since this is the region where the difference between the sizes becomes significant. 
This highlights both the cases of constant LRoM and decaying LRoM with increasing size.
These observations demand that the scaling relation of the magic be calculated at the QCP with the size of the system, though at a finite temperature approximation of the ground state considered here, the critical point might be different from $h=1$.
We performed a power law fit with equation $R-R_0 = aL^{-b}$, with $R=$LRoM. The parameter $R_0$ is the infinite system size limit (offset) of magic, $b$ provides the power-law exponent characterizing magic with the size of the system at the QCP, and $a$ is simply the constant factor. For the $2+2$ subsystem, the exponent for LRoM is found to be $0.24\pm 0.01$, whereas for the $4+4$ subsystem, the exponent is $0.34\pm0.02$. We also observed that LRoM goes to the finite offset value for the large-$L$ limit. These results are presented in insets of respective plots in Fig.~\ref{fig:lrom_size}.

\begin{figure}[ht]
    \centering
    \includegraphics[width=8cm]{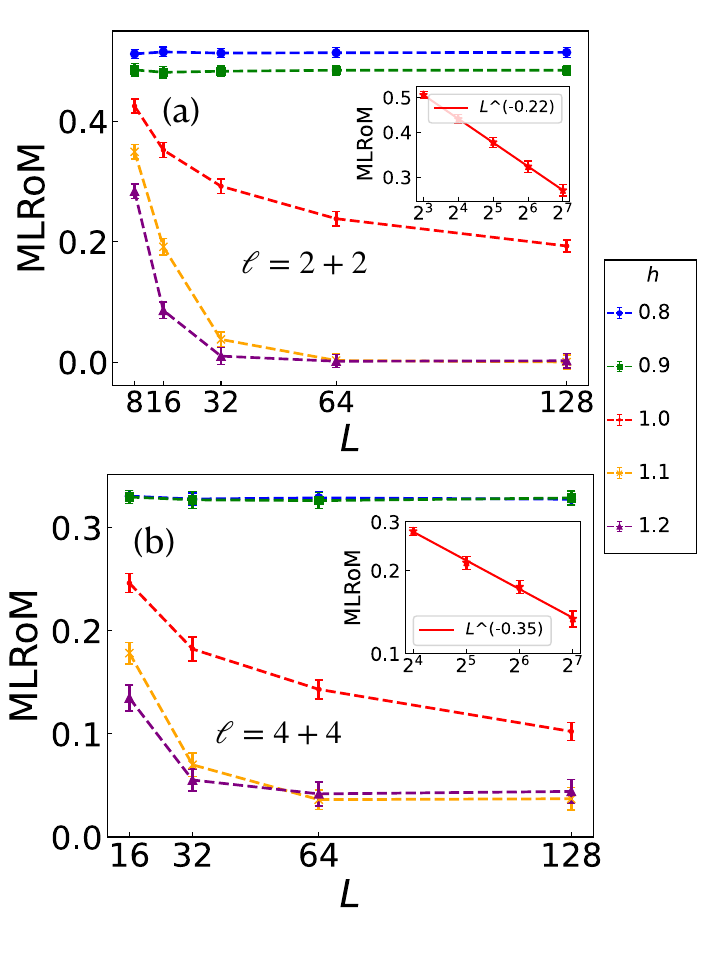}
    \caption{MLRoM with system size for different values of $h$ for the partition size of (a) $2+2$ and (b) $4+4$. The power law scaling at the critical point $h=1$ is presented in the insets.}
    \label{fig:mlrom_size}
\end{figure}

Similar size dependence is also studied for the MLRoM, for which we also observed both the constant and decaying behavior with the total size of the system. The results are plotted in Fig.~\ref{fig:mlrom_size}(a) for $2+2$ and Fig.~\ref{fig:mlrom_size}(b) for $4+4$ subsystem along with the power law behavior at the critical point reported in the insets, where scaling is performed with the same function as above with $R=$MLRoM. The exponents for MLRoM are $0.22\pm 0.02$ and $0.35\pm0.01$ for the $2+2$ and $4+4$ subsystems, respectively. In this case, the offset term decays to zero for large system sizes. 

Whereas the exponent values for LRoM and MLRoM are the same for equal subsystem sizes, they differ for different subsystems. The power law behavior is present at the critical point, but the physical setup considered here for RoM calculation alone is not able to relate to other critical properties and scaling dimensions for the 1D TFIM.
In the meantime, these exponents are smaller than the power law decay exponents of the respective four-point and eight-point connected correlation functions at the critical point calculated by Wick's theorem using the method described in Ref.~\cite{bigan2024quantumising}.

\subsubsection{Temperature Dependence}
\begin{figure}[htbp]
    \centering
    \includegraphics[width=8cm]{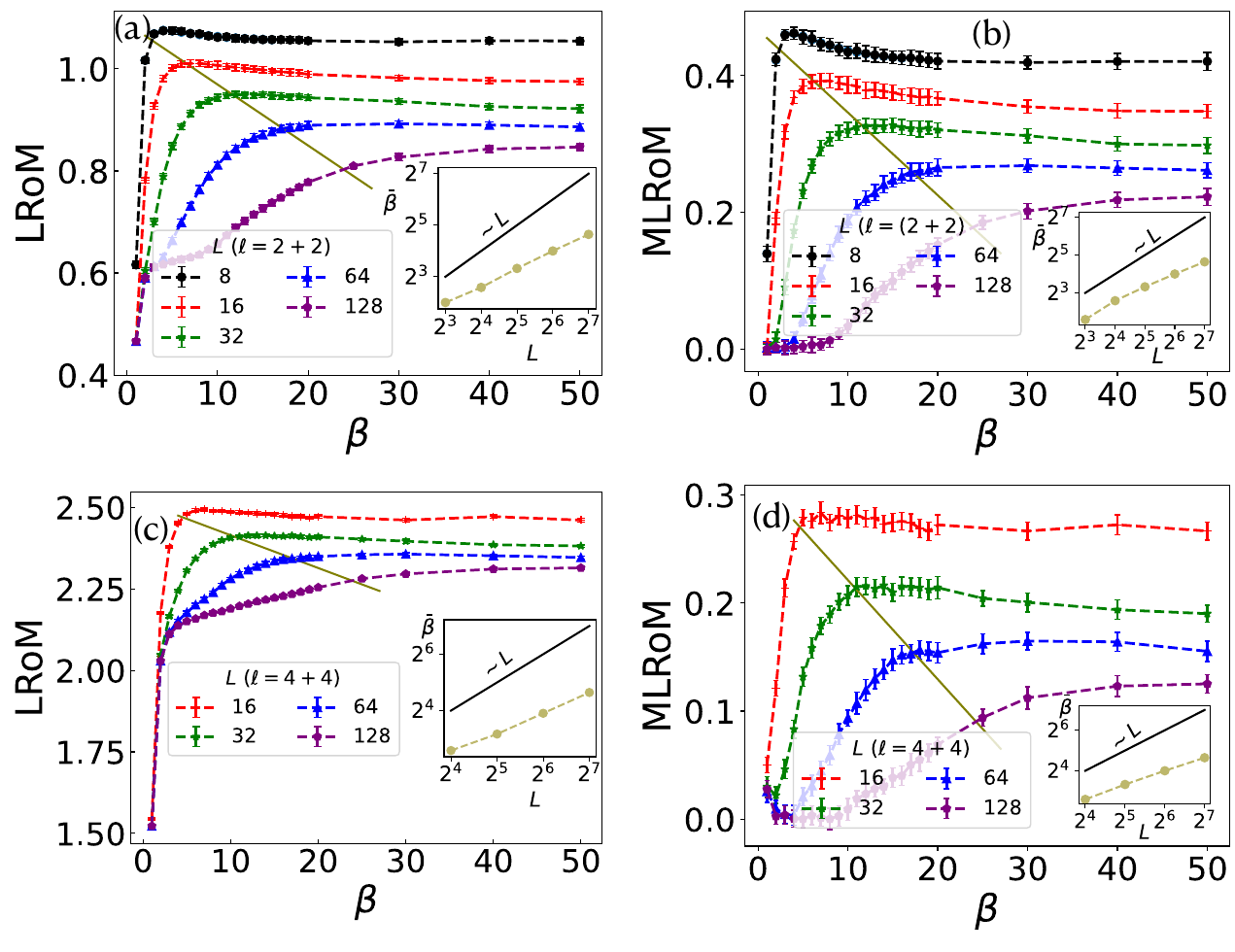}
    \caption{ LRoM ((a) size 2+2 and (c) 4+4) and MLRoM ((b) size 2+2 and (d) 4+4) with $\beta$ at $h=1$. We extract the $\bar{\beta}$ to approximate the low temperature behavior, the size dependence of which is shown in the inset with a dashed line. The relation is seen to follow the power law $(\bar{\beta}\sim L^{-z})$ with exponents $z\sim 0.67$ for LRoM and $z\sim 0.75$ for MLRoM, for both partition sizes. The solid line in the insets is the $z=1$ value, which we expect in the case of correlations for the Ising criticality. 
    }
    \label{fig:betabar}
\end{figure}

Now we see the variation of RoM with (inverse) temperature $(\beta)$, restricting ourselves to the QCP $h=1$ and for the same system sizes as for the field dependence. 
The RoM results are plotted for both subsystems, with Figs. \ref{fig:betabar}(a) and \ref{fig:betabar}(b) showing the $2+2$ LRoM and MLRoM, respectively, and Figs. \ref{fig:betabar}(c) and \ref{fig:betabar}(d) displaying the $4+4$ LRoM and MLRoM.
Initially, both the LRoM and MLRoM values increase from the lowest value to a constant value at a finite temperature. This increasing rate is higher for the smaller system sizes, and the saturation point goes towards lower temperatures for larger system sizes. 
A similar observation for the size dependence of the effective ground state for Fig. \ref{fig:romvsh} holds here also for all temperature ranges.
The stable plateau after the finite $\beta$ is similar for the full system (see Fig. \ref{fig:romfull}(a)), but the size dependence is important here as the plateau starts at different values of $\beta$ for different $L$. 
This triggered us to find the value $\bar{\beta}$, which approximates the region that separates changing RoM from the saturated one. 
By the similar nature of all plots, this point can be identified as the inflection point in the graph. But due to a limited number of points to reliably perform the finite differences, we use an alternative threshold-based approach. 
The point is determined as the value of $\beta$ where the magic is found to differ by more than a certain threshold from the low-temperature value. We choose the low-temperature value as the rightmost point of the respective curves and the threshold to be 15\%.  The approximation of which is also presented in Fig.\ref{fig:betabar} by a solid line cutting the curves. It is also evident that the $\bar{\beta}$ point depends on the system size, with the monotonic increasing relation clear from the inset in the same figure.

The results on the LRoM also show that it is nonzero for any $\beta>0$, thus demonstrating a resilience of magic even at high temperature. This is in contrast to entanglement negativity, a measure of mixed-state entanglement, which is known to vanish beyond a certain ``sudden death" temperature. This implies that there are temperature regimes where, despite the state exhibiting a positive partial transpose (thus, entanglement is not distillable), the state is nonetheless magical.

\section{Conclusions} \label{sec:conclusion}

We have presented an algorithm that, by combining quantum Monte Carlo simulations with column generation methods, allows for practical computation of the robustness of magic of partitions up to eight spins, embedded in large many-body quantum systems. This framework allows for to computation of a rigorous measure of magic for mixed states in spin-1/2 systems, going beyond the case of partitions made of single spins. 

Building on the aforementioned method, we have carried out an extensive numerical investigation of the robustness of magic in the context of the Ising chain. To better enable a physical interpretation of our results, we have introduced a new monotone, which we dubbed Log-free robustness of magic (LRoM), which satisfies subadditivity.  Similar to stabilizer R\'enyi entropies, the LRoM of the full ground state displays a maximum that lies within $10\%$ of the position of the critical point even at small volumes ($L\leq 8$) that are computationally viable. At finite temperature instead, LRoM stays approximately constant up to a value of $\beta$ that is roughly volume independent. Both of these results, while qualitatively reflecting the sensitivity of magic to phase transitions, do not bear any clear signature of criticality.

A clear signature of critical behavior is instead provided by the tri-partitioning of the system into $A_1, A_2$, and $B$ - the latter being a disconnected partition. Both LRoM and the mutual LRoM between $A_1$ and $A_2$ feature strong dependence on partition size, volume, and strength of the magnetic field. For small partition size, both quantities peak away from the critical point: this behavior is akin to that of concurrence, and might be just a consequence of the fact of an underlying monogamy of magic. For a generic partition size, the decay of mutual LRoM is exponential as a function of the distance between partitions, with the exception of the critical point: there, we observe a clear power law decay, with an exponent that is size dependent, and that does not directly match any straightforward combination of correlation functions. The emergence of a critical behavior for magic survives up to temperatures that vanish algebraically with system size.

The behavior of genuine measures of magic for mixed states remains largely unexplored for spin systems: our work represents a first attempt to go beyond the case of single sites, already considered in Ref.~\cite{sarkar2020characterization}. Importantly, the methodology we develop is generic and can be immediately implemented in other Monte Carlo schemes~\cite{tarabunga2025bell} and adapted to tensor network wave functions. This flexibility shall allow for investigations in a variety of settings, from other instances of strongly correlated matter to circuit dynamics~\cite{turkeshi2025magic,leone2025noncliffordcostrandomunitaries,magni2025quantum,PhysRevA.108.042408,fux2024disentangling,PRXQuantum.5.030332}, Hamiltonian dynamics~\cite{PhysRevA.108.042407,tirrito2024anticoncentration,odavic2024stabilizer,falcao2025magic,tirrito2025universal}, and might potentially be useful to connect to other instances of magic correlations~\cite{cao2024gravitational,korbany2025long,wei2025long}. From the experimental side, it might be interesting to prove the mutual robustness we introduced in the context of particle physics experiments, where the measurements of mutual SREs have recently been reported.

\begin{acknowledgments}
We thank M. Frau and R. Nehra for insightful discussions and feedback on the manuscript. 

M.D. is partly supported by the QUANTERA DYNAMITE PCI2022-132919, by the EU-Flagship programme Pasquans2, the PRIN programme (project CoQuS), and the ERC Consolidator grant WaveNets. M.C. and M.D. are partly supported by the PNRR MUR project PE0000023-NQSTI.
Y.M.D. and Z.Y. are supported by the Scientific Research Project (No. WU2024B027) and the Start-up Funding of Westlake University. P.S.T. acknowledges funding by the Deutsche Forschungsgemeinschaft (DFG, German Research Foundation) under Germany’s Excellence Strategy – EXC-2111 – 390814868. B.B.M. acknowledges the Natural Science Foundation of Shandong Province, China (Grant No. ZR2024QA194). The authors also acknowledge the HPC Centre of Westlake University and Beijing PARATERA Tech Co., Ltd., for providing HPC resources. 
E.\,T. acknowledges support from  ERC under grant agreement n.101053159 (RAVE), and CINECA (Consorzio Interuniversitario per il Calcolo Automatico) award, under the ISCRA initiative and Leonardo early access program, for the availability of high-performance computing resources and support.

H.T. and Y.M.D. contributed equally to this work.

\end{acknowledgments}

\appendix

\section{\label{appendix:rdm} Sampling a Reduced Density Matrix}
The thermal density matrix $\rho$ at inverse temperature $\beta$ can be written in terms of the normalized Boltzmann weight, i.e. $\rho = \exp(-\beta H)/Z$, where $H$ is Hamiltonian of the system and $Z = \Tr [\exp(-\beta H)]$ is the partition function. 
The calculation of $\rho$ involves a full diagonalization of the generic Hamiltonian, which grows exponentially with the size of the system and is limited to a smaller number of lattice sites, usually $L \leq 14$. 
The more interesting quantity in quantum information and many-body theory perspective is the RDM obtained from the full density matrix by removing the subsystems other than the required one. This can be done with partial tracing of the environment degrees of freedom $B$ from the combined system $AB$: $\rho_A={\rm Tr}_B(\rho_{AB})$, the trace being carried out over the entire degree of freedom of $B$. 
In other words, the RDM can also be treated as the effective thermal density matrix of the entangling Hamiltonian $H_E$: $\rho_A= \exp(-\beta H_E)$. The calculation of $\rho_A$ through ED is again restricted to larger systems by the complexity of $\rho_{AB}$. Therefore, we have to use some other efficient numerical techniques to push the boundaries to larger system sizes. 
To this end, we will explain the techniques provided by the QMC+ED method presented in Ref.~\cite{mao2025sampling}. The QMC is done via the path integral QMC with finite temperature stochastic series expansion (SSE). 
This algorithm is efficient in the sense that its complexity is polynomial with respect to the total size of the system; however, it is still limited by the reduced system size, whose exponential complexity cannot be eliminated due to the size of the RDM matrix.

We first introduce the standard SSE method, then explain how to modify it to sample the RDM of interest.
The detailed implementation of SSE depends on the models of interest. 
Typically, we write the Hamiltonian as the sum of local operators $\{H_{a,b}\}$
\begin{equation}
    H = -\sum_{a,b}^L H_{a,b},
\end{equation}
where $a$ denotes the type (diagonal or off-diagonal) of the local operator in a suitably chosen basis, and $b$ denotes the spatial degree of freedom (e.g., a site or a bond).
The partition function $Z={\rm Tr}\lbrace {\exp}(-\beta H)\rbrace$ is then expanded in a power series, with the trace expressed as a sum over diagonal and off-diagonal matrix elements, which is 
\begin{equation} \label{zn}
    \begin{split}
    Z &= \sum_{\alpha} \sum_{n=0}^\infty \sum_{S_n}\frac{\beta ^n}  {n!}\left\langle \alpha \right| \prod_{l=1}^n H_{a(l),b(l)} \left| \alpha \right\rangle,
    \end{split}
\end{equation} 
where $S_n$ is a sequence of $n$ operator-index pairs
\begin{equation}
	S_n = [a(1),b(1)],\ldots ,[a(n),b(n)].
\end{equation}
For TFIM, we consider the standard $\sigma^z$ basis.
By introducing a sufficiently large cut-off $\Lambda$ to the series expansion in Eq.~\eqref{zn} such that the truncation error is negligible, we can construct an efficient sampling scheme. 
The length of each operator sequence is kept constant by inserting a null operator $\Lambda-n$ times in the product in \eqref{zn}, and there are $\binom{\Lambda}{n}$ ways for each insertion, yielding the partition function
\begin{equation} \label{zl}
    Z = \sum_{\alpha} \sum_{S_M} \frac{\beta^n (\Lambda-n)!}{\Lambda!} \left \langle \alpha \right | \prod_{l=1}^{\Lambda} H_{a(l),b(l)} \left | \alpha \right \rangle ,
\end{equation}
where we denote the null operators by $[a(l),b(l)] = [-1,-1]$, and $n$ gives the number of non-null operators. 

For the TFIM, we usually introduce a constant shift to facilitate the simulation. To this end, we rewrite the Hamiltonian of TFIM in the form
\begin{equation}
    H = -J\sum_{\langle ij\rangle}^L (\sigma_i^z\sigma_{j}^z + I) - h \sum_{i=1}^L(\sigma_i^x + I).
\end{equation}
Here, the first sum is over lattice bonds, while the second sum is over lattice sites. Next, we decompose the lattice  Hamiltonian into the bond and site operators~\cite{sandvik2003ising}
\begin{align*}
    & H_{0,i} = hI_i,\\
    &H_{1,i}= h\sigma_i^x \\
   &H_{2,ij} = J(\sigma_i^z \sigma_j^z+I).
\end{align*}
If $|\bullet\rangle$ and $|\circ\rangle$ be the $+1$ and $-1$ eigenstates of $\sigma^z$, then all the nonzero matrix elements are positive
\begin{align*}
    &\langle\bullet |H_{0,i}|\bullet \rangle =     \langle\circ |H_{0,i}|\circ \rangle = h, \\
    &\langle\bullet |H_{1,i}|\circ \rangle =     \langle\circ |H_{1,i}|\bullet \rangle = h, \\
    &\langle\bullet\bullet |H_{2,ij}|\bullet\bullet \rangle = \langle\circ\circ |H_{2,ij}|\circ\circ \rangle =2J,
\end{align*}
and can be used as relative probabilities satisfying the detailed balance condition in a sampling procedure, therefore forming the basis of the SSE representation for the TFIM.
Now we define the MC updating procedure for both diagonal and off-diagonal operators.

\textit{Diagonal update:}
We traverse through the list of all $\Lambda$ operators sequentially in the propagation. 
If the off-diagonal operator $H_{1,i}$ is encountered, it is ignored, but the $\sigma^z$ spin is flipped at that site.
If the diagonal operator ($H_{0, i}$ or $H_{2,ij}$) is encountered, it is removed with a certain probability that can be predefined. 
If  $H_{-1,-1}$ is encountered, we choose to insert a diagonal operator of the type $H_{0, i}$ ($h$-term) and $H_{2,ij}$ ($J$-term) with probabilities
\begin{equation}
    P(h) = \frac{hN}{hN+(2J)N_b},\quad P(J) = 1-P(h),
\end{equation}
and accept the addition of an operator with probability
\begin{equation}
    P = \text{min}\left(\frac{\beta(hL+(2J)N_b)}{\Lambda-n},1  \right)
\end{equation}
to insert into the randomly chosen appropriate bond or site.
Each diagonal update modifies the operator positions, and the topology of the cluster structure is fixed after each update.

\textit{Cluster update:} 
These are non-local updates capable of large-scale changes to the configuration of the simulation cell and ensure the ergodicity in sampling of the TFIM model.
A cluster is formed in the $1+1$ simulation cell by grouping spins and bond operators according to: clusters terminate on site-operators $H_{0, i}$ or $H_{1,i}$, and bond operators $H_{2,ij}$ belong to one cluster.  
This procedure continues until all clusters are identified.
Notice that the switching between diagonal $H_{0, i}$ and off-diagonal $H_{1,i}$ operators involves no weight change since they have the same matrix elements. 
The Swendsen-Wang procedure flips clusters, simultaneously sampling the site operator type and the spin configuration $|\alpha\rangle$.

Next, we introduce the modifications.
To sample RDM $\rho_A=\Tr_{B}(\rho_{AB})$, we employ an open boundary condition for subsystem $A$ in the temporal axis, compared with simulating $Z$. The diagonal and off-diagonal updates can then be performed similarly. In this case, an SSE configuration has different initial and final states $\ket{C_A}$ and $\ket{C_A'}$, for subsystem $A$. The frequency of sampling such configurations is therefore proportional to the matrix element $(\rho_{A})_{C_A,C_A'}$. By collecting a sufficiently large number of samples, we can reconstruct the RDM matrix $\rho_A$.

\section{\label{appendix_CG} Calculation of RoM}
\textit{Naive method}: The naive method involves the minimization over the solution of linear equations obtained from the matrix equation $\mathbf{A}_n\mathbf{x}=\mathbf{b}$. Since the dimensions of $\mathbf{A}_n,\mathbf{x},\mathbf{b}$ are $4^n\times|\mathcal{S}|_n$, $|\mathcal{S}|_n\times 1$, $4^n\times 1$, we have to find the suitable set of $|\mathcal{S}|_n$ parameters satisfying $4^n$ constrains. Many parameters may be zero, but the trivial restriction is to ensure the affine combination $\sum_i x_i=1$. For pure stabilizer states, the only solution is for one parameter to be unity while the rest should vanish. Also, for mixed stabilizer states, we can find the set of optimal parameters such that all are nonnegative, which is easily seen from the definition in Eq.~\eqref{eq:stabn}. Therefore, for the stabilizer states $\text{RoM} = 1$.  For nonstabilizer states, there should be at least one negative value of $x_i$ and $\sum_i|x_i|>1$, which says that the RoM is greater than one. 

\textit{Column Generation}: The CG method \cite{desaulniers2006column,hamaguchi2024handbook} is an iterative procedure of computation and utilizes the dual formalism of RoM defined in Eq.~\eqref{eq:romdual}. 
The CG method is the modified version of the random selection and top-overlap methods presented in Ref.~\cite{hamaguchi2024handbook}, together with the construction of $\mathbf{A}_n$ matrix and the calculation of stabilizer overlaps $\mathbf{A}_n^\top \mathbf{b}$. 
The top-overlap method restricts the number of columns in $\mathbf{A}_n$ by considering only a fraction $K$ (ensuring $K|\mathcal{S}|_n>4^n$) of stabilizer states with the largest or smallest overlaps (the other fraction $(1-K)$ is neglected) and calculates the RoM using the definition of Eq.~\eqref{eq:rom1}.  
The idea of the CG method is to start the calculation of RoM with suitably chosen $K$ fractions of the columns of $\mathbf{A}_n$ by the top-overlap method (denote the reduced columns set by $\mathcal{C}_k$) and to check the inequality of overlaps $ |\mathbf{a}^\top\mathbf{y}|\leq 1,\, \forall \mathbf{a}\in \mathbf{A}_n$ in terms of the dual variable $\mathbf{y}$. At each iteration, the inequality-violating columns are identified and added to update the initial choice and proceed with the calculation using the new $\mathcal{C}_k$ until all constraints are satisfied.
To be consistent, we add only up to $K$ fractions at each iteration step.
While the naive calculations are limited to 5 qubits, we can go up to 8 qubits by using the CG technique regardless of the nature of the (reduced) density matrix. 
The choice of pairs of the number of qubits and initial fraction $(n, K)$ is as follows: $(5,10^{-1}),(6,10^{-3}),(8,10^{-8})$. 
The CG method can also be applied to compute other nonstabilizerness measures involving the $\mathbf{A}_n$ matrix, for example, the stabilizer fidelity and the stabilizer extent~\cite{Bravyi2019simulationofquantum,hamaguchi2025faster}.
We can implement these methods of convex optimization routine using CVXPY~\cite{diamond2016cvxpy}, Gurobi~\cite{gurobi}, or any equivalent solvers. 

\bibliography{references}

\end{document}